# Game-Theoretic Model and Experimental Investigation of Cyber Wargaming

Edward J. M. Colbert
*US Army Research Laboratory*
Adelphi, USA
edward.j.m.colbert@gmail.edu

Alexander Kott
*US Army Research Laboratory*
Adelphi, USA
alexander.kott1.civ@mail.mil

Lawrence P. Knachel, III
*US Army Communications-Electronics Research, Development and Engineering Center*
Adelphi, USA
lawrence.p.knachel2.civ@mail.mil

*Abstract*— We demonstrate that game-theoretic calculations serve as a useful tool for assisting cyber wargaming teams in identifying useful strategies. We note a significant similarity between formulating cyber wargaming strategies and the methodology known in military practice as Course of Action (COA) generation. For scenarios in which the attacker must penetrate multiple layers in a defense-in-depth security configuration, an accounting of attacker and defender costs and penetration probabilities provides cost-utility payoff matrices and penetration probability matrices. These can be used as decision tools by both the defender and attacker. Inspection of the matrices allows players to deduce preferred strategies (or COAs) based on game-theoretical equilibrium solutions. The matrices also help in analyzing anticipated effects of potential human-based choices of wargame strategies and counter-strategies. We describe a mathematical game-theoretic formalism and offer detailed analysis of a table-top cyber wargame executed at the US Army Research Laboratory. Our analysis shows how game-theoretical calculations can provide an effective tool for decision-making during cyber wargames.

## I. Introduction and Motivation

Cyber wargaming is increasingly often used by commercial and public-sector organizations. Such a wargame typically involves actual employees of an organization who play roles in a human-based (and occasionally computer-based) simulation of a cyber attack and responses to the attack. A number of consulting organizations provide wargaming design and facilitation as a service [1]. The growing prominence of cyber wargames is hardly surprising. Cyber conflicts involve problems of adversarial nature, not unlike those in military practice, and these are often solved by resorting to game-theoretic models, or by simulations – wargames.

However, theoretical foundations and guidance for cyber wargaming are lacking. In this paper, we offer a game-theoretic model of cyber attack and defense, compare the model-based analysis with experiences of an actual table-top wargame, and offer recommendations on using game-theoretic analysis for enhancing accuracy and value of such cyber wargames.

### A. Current Practices of Cyber Wargaming

In spite of its growing popularity among corporate and government organizations [2], the term "cyber wargame" is not particularly well defined and may refer to many different forms of an exercise, test, simulation or emulation event. Typically, unlike penetration testing in which "white hat" hackers seek to find the company's technical vulnerabilities, a corporate cyber wargame often places emphasis on a business scenario involving a cross-section of the company's business functions [3,4,5]. Modern cyber wargaming in corporate and government scenarios [6,7] utilize the insights and experience gained from military wargaming [8,9,10].

The wargame is structured – often by a specialized consulting organization hired for this purpose [1] – to simulate experiences of a real cyber attack, and realistic responses to it. Participants of the wargame often comprise the company's employees from multiple functional areas: information security, application development, network operations, facilities management, customer service, production, marketing, legal and public affairs, financial, and distribution.. These players gather in one or several conference rooms, for a duration of anywhere from 4 hours to 3 days, and under the guidance of professional facilitators proceed to enact the events of a cyber attack, usually developing a strong commitment and passion in the game.

Great diversity in the types and forms of cyber wargames is found in current cyber exercises. Such diversity can be characterized along several dimensions:
- Breadth of business functions: the focus of a wargame may range from strictly technical considerations of vulnerabilities, capabilities and activities of software and hardware, to a broad coverage of business functions, e.g., financial, media, legal and business operations aspects, where the technical cyber compromise is merely a starting point of the scenario. This may correlate with the seniority of decision-makers involved: broader scenarios may involve leaders higher on the corporate ladder.
- Scale of an entity under consideration: a wargame may concern itself with an entity limited to a single web server to large-scale operations or a multi-national corporation; it could be a single system or network, a site or an enterprise or an international system of enterprises.
- Realism of the game: a wargame may range from a tabletop exercise with little more than paper and pencil; to computer-assisted simulations; to use of emulated cyber ranges; and even to attacks on operational systems.

The process of designing, implementing and executing a cyber game normally involves most of the following steps (not necessarily sequential or in this order). They could be performed by a consulting organization in close collaboration with personnel of the company [2]:
- Defenses: Identify security mechanisms, tools and personnel; their attributes and capabilities.
- Threats: Hypothesize suitable threats, their capabilities and likely TTPs (Tactics, Techniques and Procedures), goals, limitations, access opportunities, skill levels, time and resources available to them.
- Attacks: Formulate and select attack scenarios, including at least two -- most likely to be executed by the threat against this organization and most dangerous – the one that may be less likely but would cause the greatest damage.
- Players: Recruit relevant participants-defenders from the company; who are relevant to the site or enterprise being wargamed; and who would realistically be engaged in defending against a given a threat scenario.
- Blue cell(s): Organize the participants-defenders into a team or teams responsible for planning and executing defensive actions; such a team is called a Blue cell; preferably, participants are organized into teams (cells) that are reflective of the actual organizational structure of the company.
- White cell: Provide a white cell, i.e., trained individuals who have experience in wargames and can document important information emerging during the wargame, guide and facilitate the game, and adjudicate or arbitrate outcomes of individual actions taken by participants. These are usually outside consultants.
- Red cell: Provide or designate a team of individuals who play the role of the attacker; such a team is called the Red cell.
- Rooms and Props: Prepare physical facilities, means of communication, paper or computer-based products to conduct the game.
- Play: Assemble all cells, begin execution of the scenario, let the Red cell attack, Blue cell defend, and White cell declare the status as it evolves and inject additional events to keep the game moving in the right direction.
- Payoff: Analyze the observations and results of the game, and formulate recommendations.

The outcomes and benefits of the cyber wargame vary depending on the goals of the organization. These may include identification of hidden vulnerabilities, incorrect assumptions, and need for additional procedures and training.

A cyber wargame can also help identify poorly understood risks; educate and entertain personnel; obtain support of senior decision-makers; explore the extent of potential disruption to various business functions; clarify the roles and responsibilities of cyber responders; improve communication among them; allow stakeholders to get to know one another, and build relationships; understand decision-making authorities; highlight interactions with third-party business partners; identify potential gaps in an organization's preparedness and response plans.

*B. Military Practice of Course of Action Analysis*

Much of the techniques used in cyber wargaming appear to be influenced or directly borrowed from military wargaming practices. The US Navy War College has provided extensive expertise and documented guidance for military wargaming [9,10]. In the US Army, wargaming practice is often called Course of Action (COA) Analysis [11,12]. Although such analysis is performed by military units of various sizes -- from a small unit called squad to a very large organization called corps – here we will use an example of a unit called Brigade Combat Team (BCT).

Somewhat comparable to a mid-size corporation, a US Army BCT includes several thousands of professional soldiers and officers, hundreds of combat and support vehicles, helicopters, sophisticated intelligence and communication equipment and specialists, artillery and missiles, engineers, medical units, and repair shops. In a battle, these assets might perform hundreds of complex tasks of multiple types (similar to corporate "business functions"): collecting intelligence; movements; direct and indirect fires; constructing roads, bridges, and obstacles; transporting and handling supplies; managing the civilian population; command and control, and so on. Unlike in cyber wargaming, the threat (i.e., the enemy that the BCT fights against) tries to apply much physical destruction to the BCT, although cyber attack are often also a part of the threat's repertoire.

Detailed planning of a military operation requires an intensive effort of highly trained professionals, called the BCT planning staff. Typically, it consists of four or five officers, ranging in rank from captain to lieutenant colonel, who perform this work with the support of a subordinate staff. The process normally takes from two to eight hours – not unlike a typical cyber wargame. The physical environment often consists of a tent extended from the back of one or several Army trucks or armored command-and-control vehicles; folding tables and chairs, and -- similarly to a cyber wargame -

- either computer screens or paper maps on which the officers draw symbols of units and arrows of movements.

The input for the staff's effort comes usually from the unit commander as a high-level specification of the operation. With this input, the planning staff works as a team – called the Blue cell – perform actual wargaming, including:

- Predicting enemy actions or reactions. This is done by the Red cell, usually the officer who specializes in collecting and analyzing enemy intelligence [13]. The Red cell plays the role of the enemy in order to help the Blue cell understand possibly actions and responses of the enemy. Similarly to the cyber wargames, Red cell provides two cases of enemy actions: the most likely plan of enemy actions, and the most dangerous (to the BCT) plan of enemy actions. The latter might be the same as the former, but usually is different as it involves assumptions of greater capabilities on the part of the enemy.
- Planning and scheduling the detailed tasks required to accomplish the specified COA [14], and to prevent or respond to the threat actions (like the Blue cell defenders would do in a cyber wargame), and allocating tasks to the diverse forces constituting the BCT (like elements of a corporate response to cyber attack).
- Estimating success of failure of friendly and enemy actions, and battle losses [15]. This is similar to the function performed by the White cell in cyber wargaming.
- The process of estimating enemy actions and friendly actions may repeat in several cycles until a convergence is achieved: the Red cell is unable to suggest any further improvements of the enemy actions, and the Blue cell is unable to suggest any further improvements of the friendly actions. This hints at reaching something akin to Nash equilibrium, in game theoretic terms.

This wargaming usually produces a plan/schedule in a synchronization matrix format, a type of Gantt chart (see Figure 1). The chart's columns represent time periods. The rows contain functional classes of actions, such as Maneuver (which in turn includes such subclasses as Main Effort and Security), Combat Service Support (for example, logistics), Military Intelligence, and so on. This plan-schedule's content, recorded largely in the matrix cells, includes the tasks and actions of the friendly force's subunits and assets, their objectives and manner of execution, expected timing, dependencies and synchronization, routes and locations, availability of supplies, combat losses, enemy situation and actions, and so on.

Ultimately, the purpose of the military wargame is for the Blue cell to consider and select a small (manageable, often on the order of three) number of COAs that are seen as most advantageous to the Blue side. In doing so, the Blue cell has to make an assumption about the COA that would be adopted by the Red side. In military wargaming, there are several ways to approach this difficult decision.

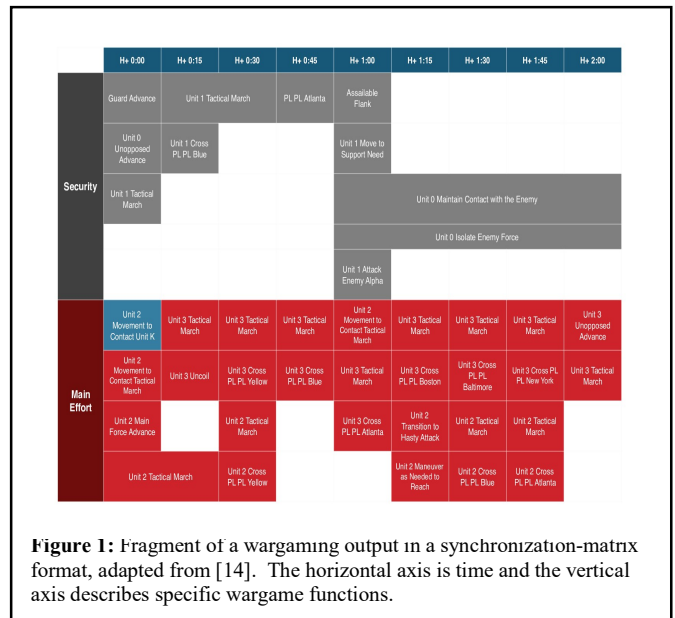

**Figure 1:** Fragment of a wargaming output in a synchronization-matrix format, adapted from [14]. The horizontal axis is time and the vertical axis describes specific wargame functions.

One approach is to consider the "most likely" COA of the opponent, i.e., the Red COA that the Blue cell feels is most likely to be adopted by the Red side. This assessment of likelihood might be based on Blue cell's knowledge of the Red side's preferences, e.g., the COAs that the Red side has adopted in previous battles. Alternatively, the Blue cell might decide that the most likely Red COA is the one that provides the Red side with the greatest advantage or greatest utility in the battle.

Another approach might be to consider the "most dangerous" or "most damaging" Red COA – the COA that would cause the greatest damage to the Blue side. Note that the "most likely" and "most damaging" Red COAs are often different.

Having selected the "most likely" Red COA, and the "most damaging" Red COA, the Blue cell usually attempts to select a Blue COA that would perform sufficiently well against both of the Red COAs.

Besides creating this tangible set of potential strategies, other valuable outcomes of this wargaming are similar to those of corporate cyber wargames: identification of hidden vulnerabilities, incorrect assumptions, risks and losses, education, clarity of roles and responsibilities. It is important to note that the execution and modeling of a wargame is not necessarily representative of a real-world encounter. Real-world actors likely do not have complete knowledge of the strategies and limitations of their opponents and a systematic or mathematical formulation for computing optimal strategies may be difficult or impractical. While complex models may be able to be constructed for real-world scenarios, it is outside the scope of this paper to do so; our simple model is mean to apply only to cyber wargaming.

In the remainder of the paper, we first propose a game-theoretic model of a contest between cyber attacker and cyber defender. Then we describe an actual cyber wargame designed and led by one of this paper's authors. We explore how our theoretical model may apply to the actual wargame. Finally, we discuss the insights and benefits that the model brings to the cyber wargame, and offer a set of practical recommendations for designing and conducting cyber wargames.

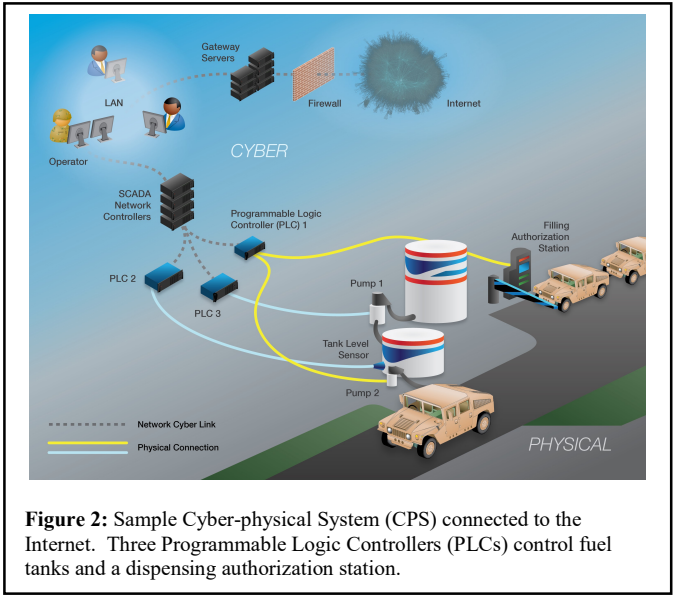

**Figure 2:** Sample Cyber-physical System (CPS) connected to the Internet. Three Programmable Logic Controllers (PLCs) control fuel tanks and a dispensing authorization station.

II. GAME-THEORETIC METHOD

In this section, we formulate an approach to analysis of a cyber wargame problem that combines elements of game theoretic and risk analytic treatments. We believe our approach is useful as a computational tool to aid in complex wargame and training exercises in which the strategy space is larger or more complex than is normally possible for most humans to explore. In a later section of the paper, we then discuss an example of a real-world process where an approximation of this approach is used.

*A. Mathematical Model*

In the following formulation, we are partly inspired by [16]. Consider a cyber-physical system in Figure 2. For this system, a cyber attacker desires to obtain a benefit $b$ by accessing the system via the Internet and eventually obtaining control of the plant's Programmable Logic Controller (PLCs). In doing so, the attacker would have to penetrate defensive mechanisms and overcome actions of the defenders in several layers of the cyber-physical system.

To make the discussion more general, in Figure 3, we illustrate an abstract notion of the problem: an attacker enters the outer attack surface with the intention of penetrating a series of layers guarded by the defender before arriving at the target. Penetration of the layers will require specific malicious actions, and the overall path to the target determines the precise attacker "strategy" that the attacker intends to use.

We describe the plans of the attacker with a set of $N_a$ potential strategies $\{s_{a,i}\} \in S_a$ where $S_a$ is the attack strategy space. The attack strategies are identified by index $i$. Likewise, the defender has developed a set of $N_d$ strategies $\{s_{d,j}\} \in S_d$ where $S_d$ is the defense strategy space and $j$ is the defense strategy index. In this work we describe a simplified model in which both attacker and defender have complete knowledge of the system and can therefore determine each other's strategies. The defender strategies are accomplished by selecting specific subsets of cyber-defense mitigations $\{m_{d,k}\} \in M_d$ where $M_d$ is the set of all mitigations, and $k$ is the mitigation index.

As shown in Figure 3, there are $N_l$ layers that the attacker needs to penetrate. We identify these layers by the index $l$.

There are costs for both the attacker and the defender for each specific strategy tuple $\{i,j\}$. That is, given an attack strategy $i$ and a defense strategy $j$, the attacker suffers a cost $C_{a,ij}$ to accomplish his goal and the defender spends a cost $C_{d,ij}$ to deploy his defense strategy.

Finally, given a strategy choice $\{i,j\}$, the attacker is assumed to penetrate layer $l$ with probability $p_l(s_{a,i}, s_{d,j})$, or $p_{l,ij}$ in shorthand notation. As shown in **Figure 3**, if the attacker penetrates all $N_l$ layers, he reaches the target $T$ and obtains a benefit $b$.

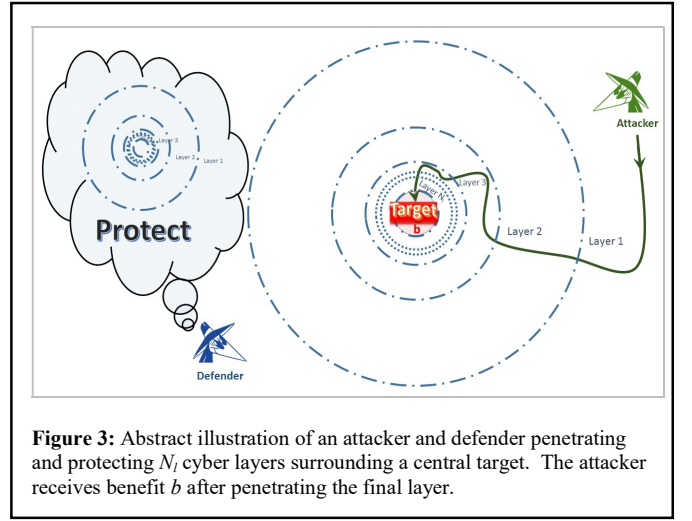

**Figure 3:** Abstract illustration of an attacker and defender penetrating and protecting $N_l$ cyber layers surrounding a central target. The attacker receives benefit $b$ after penetrating the final layer.

In the model below, we consider the scenario of when the attacker gains benefit $b$, the defender loses an equivalent value ($b$) of his assets. This constraint could easily be modified as the situation demands.

The attacker and defender expected utility $u_a$ depends on the benefit $b$, the total probability of penetrating all layers $P_{ij}^T = \prod_{l=1}^{N_l} p_{l,ij}$, and expended costs $C_{a,ij}$. For the attacker:

(1a) $u_a(s_{a,i}, s_{d,j}) = b \prod_{l=1}^{N_l} p_l(s_{a,i}, s_{d,j}) - C_a(s_{a,i}, s_{d,j})$,

or, in shorthand notation

(1b) $u_{a,ij} = b P_{ij}^T - C_{a,ij}$

Likewise, for the defender:

(2a) $$u_d(s_{a,i}, s_{d,j}) = b[1 - \prod_{l=1}^{N_l} p_l(s_{a,i}, s_{d,j})] - C_d(s_{a,i}, s_{d,j})$$

, or, in shorthand notation

(2b) $$u_{d,ij} = b(1 - P_{ij}^T) - C_{d,ij}$$

The challenge for each player is to select the strategy, $s_a^*$ and $s_d^*$, for attacker and defender respectively. As will be described in section II.B, there are a number of methods the attacker and defender may use to select their strategies $s_a^*$ and $s_d^*$. Upon selection of the player strategies $s_a^*$ and $s_d^*$, the total probability of penetration can be computed:

(3) $$P_T^* = \prod_{l=1}^{N_l} p_l(s_a^*, s_d^*).$$

### B. Strategy Selection

Once the attacker and defender costs and penetration probabilities are known or assumed, payoff matrices $u_{a,ij}$ and $u_{d,ij}$ can be computed and inspected by each player to determine their preferred strategy. In this section, we describe four methods by which the defenders and attackers in the wargame may choose their strategy. We discuss these methods in more detail in section IV, in the context of the specific case study outlined in section III.

#### Pure Strategy Equilibria

We first describe the case in which the players choose a single unique strategy (pure strategy approach) as opposed to estimating a probabilistically weighted set of multiple strategies (a.k.a. mixed strategy approach).

The model outlined in section II.A does not describe a zero-sum game (defined as $u_{a,ij} = -u_{d,ij}$) and a Nash equilibrium state for $s_a^*$ and $s_d^*$ may not exist. However, since the number of strategies considered will be low enough to be manageable by a human, manual methods can be used to search for a local saddle point or a Nash equilibrium.

That is, if we search over each defense strategy $j$ for the preferred attack strategies $\{i_j\}$ of the attacker

(4a) $$\{i_j\} = s_{a,i}^{max,j} = \underset{s_{d,j} \in S_d}{\operatorname{argmax}} u_a(s_{a,i}, s_{d,j}),$$

and, likewise, search over each attack strategy $i$ for the preferred defense strategies $\{j_i\}$ of the defender

(4b) $$\{j_i\} = s_{d,j}^{max,i} = \underset{s_{d,j} \in S_d}{\operatorname{argmax}} u_d(s_{a,i}, s_{d,j}),$$

and find saddle points in the payoff matrices, one or more "most likely" equilibrium strategies may be found. If only one equilibrium point is found, it is by definition a Nash equilibrium. We illustrate this method further in sections III and IV.

Equilibrium strategies such as these may be preferred by the players in the case when the payoff matrices are fully disclosed to both players and both players must choose their strategy simultaneously.

#### Strong Stackelberg Strategy Equilibria

A common method for finding solutions to non zero-sum games is assume that one of the players has the opportunity to be first in selecting his strategy, and chooses a mixed strategy solution and the opponent follows with a pure strategy (i.e. a single strategy with 100% probability). The equilibrium strategies in this scenario are known as Strong Stackelberg Equilibria (SSE) [17,18]. If the wargame is played in this manner, the leader and the follower could use one of the many methods for finding SSE solutions, such as those described in [19]. While we do not discuss SSE equilibria further in this work, we later consider an example where multiple strategies are considered to some extent.

#### Playing Against the Most Likely Strategy of the Opponent

As described in section I.B, a player may choose a strategy by first estimating the "most likely" strategy of the opponent, based on any available information about the opponent. Then the player selects his own strategy based on how successful it will be against the "most likely" strategy of the opponent.

#### Playing Against the Most Damaging Strategy of the Opponent

Alternatively, as described in section I.B, a player may choose a strategy by first estimating the "most damaging" strategy of the opponent, i.e., the strategy in which the opponent would impose the most severe losses on the player. Then the player selects his own strategy based on how successful it will be against the "most damaging" strategy of the opponent.

### C. Sample Calculation

To illustrate, we provide sample calculations for a simple scenario. Suppose a freelancing cyber-crime group is engaged by an anonymous third party to penetrate controls of a munitions plant. Two layers of the network need to be defeated: $l = 1$ and $l = 2$. If successful, the attacker will be paid a benefit $b$ of $50k. The defender of the plant consistently uses a single strategy $s_d^* = s_{d,1}$.

Based on the preliminary reconnaissance of the plant's network, the attacker considers two possible strategies. The first strategy $s_{a,1}$ will use an available exploit that will rapidly defeat the defenses of layer 1 with probability $p_{1,11} = 0.9$. However the exploit is noisy and will likely alert the defenders, thereby reducing the probability of penetrating layer 2 to $p_{2,11} = 0.1$. The cost associated with this "fast and noisy" strategy is only $C_{a,11} = \$5k$. The second strategy $s_{a,2}$ will require the development of a new, stealthy exploit for layer 1, and will take longer to deploy. In this case, $p_{1,21} = 0.9$, $p_{2,21} = 0.8$, and the costs to the attacker are much higher $C_{a,21} = \$15k$.

Using Equation (1), the expected attacker utilities for the first and second attacks are, respectively:

$$u_{a,11} = \$50k\,(0.9 \cdot 0.1) - \$5k = -\$0.5k$$

$$u_{a,21} = \$50k\,(0.9 \cdot 0.8) - \$15k = \$21k.$$

Clearly, the second attack has higher utility and therefore the attacker selects $s_a^* = s_{a,2}$. The attacker's probability of success $P_T^* = P_{21}^T = 0.9 \cdot 0.8 = 0.72$

*D. Practical Considerations*

To generalize, the overall process for calculating the game-theoretic quantities in section II.A is:
1. Collect information about $S_a, S_d, p_l(s_{a,i}, s_{d,j}), C_a(s_{a,i}, s_{d,j})$, and $C_d(s_{a,i}, s_{d,j})$ from empirical and experimental sources
2. Compute cost-utility functions (payoff matrices) and total penetration probability matrix using Equations (1), (2), and (3)
3. Utilize the payoff matrices to determine the best strategy (see section II.B)

The most difficult step may be the first one. Quantitative data, such as $p_{1,11} = 0.9$ and $p_{2,11} = 0.8$ in our illustrative example, come from subject matter experts and may be difficult to estimate. Obtaining such information, particularly from subject matter expects, may be a difficult and expensive process that can yield subjective, inconsistent and/or unreliable results.

*E. Relation to Military Wargaming Practices*

As described in section I.B, the practice and process of military wargaming (COA analysis) is as follows. Having collected and considered the relevant information, the Blue cell officers propose a friendly defense strategy $s_{d,1}^H$, where the superscript H indicates a strategy devised by a human in the military wargame. The strategy $s_{d,1}^H$ denotes a sequence of activities and associated resources, time periods and spatial locations. This strategy is recorded in the synchronization matrix (Figure 1). Then the Red cell who plays the role of the enemy proposes enemy strategy $s_{a,1}^H$. Given $s_{a,1}^H$, the Blue cell produces a modified strategy $s_{d,2}^H$, and so on. In each iteration, and for each activity, the planning team uses its experience and doctrinal guides to determine whether the activity will succeed, and how much losses each side will encounter as the result of the activity. These iterations continue until equilibrium is reached where the Red cell is unable to suggest any further improvements to their attack strategy, and the Blue cell is unable to suggest any further improvements to their defense strategy, or both cells are exhausted in an unsuccessful attempt to find such an equilibrium. Implicitly or explicitly, the losses are expected to be kept below some maximum allowable value. In the human wargame activity, risk is often assessed qualitatively by the players by estimating the likelihood of accomplishing the military objective of the battle, e.g., capturing certain terrain or destroying the enemy's forces.

In a comparable cyber wargame, our game-theoretic model and strategy selection methods offer an analytical tool for the Blue and Red cell to decide on their initial strategies $s_{d,1}^H$ and $s_{a,1}^H$. The payoff and penetration matrices can be used as tools for considering subsequent strategies as long as no targets in the layers have been compromised and the costs are not changed. While we do not offer a mathematical framework for determining subsequent strategies in the wargame where targets in the layers have been compromised and costs need to be modified dynamically, such a framework could be constructed using the same methodology.

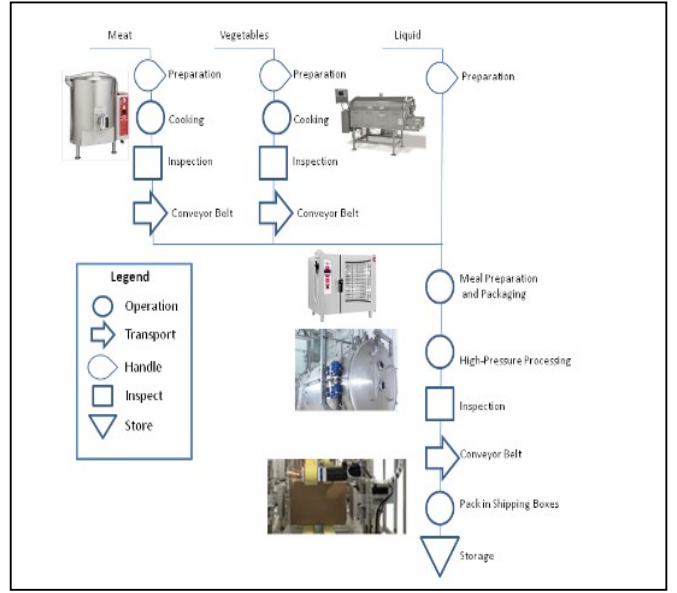

**Figure 4:** Process map for the production line in the fictitious AQUA plant

III. EXPERIMENTAL INVESTIGATION OF CYBER-WARGAMING

*A. General*

As a more elaborate example, we consider the application of our game-theoretical framework to a table-top wargaming activity conducted at the US Army Research Laboratory [20]. In this event, a fictitious AQUA cyber-physical control system (specifically, an Industrial Control System [ICS]) was designed and presented to two teams of human cyber experts – a RED team (i.e., Red cell) and a BLUE team (i.e. Blue cell) -- representing the attacker and defender, respectively.

*B. AQUA Wargame Information Packet*

A technical information packet [21] describing the AQUA ICS was provided to both teams before the exercise began. Highlights from the information packet follow.

The AQUA ICS is a food processing plant that produces packaged meals. The process map for our fictitious AQUA plant is shown in Figure 4. The plant executes six manufacturing processes. The meat and vegetables are cooked separately. Once they are cooked, the meals are prepared and packaged in a material suitable for high-pressure processing. Once the high-pressure process is completed, the meals are placed in boxes and stored in a warehouse.

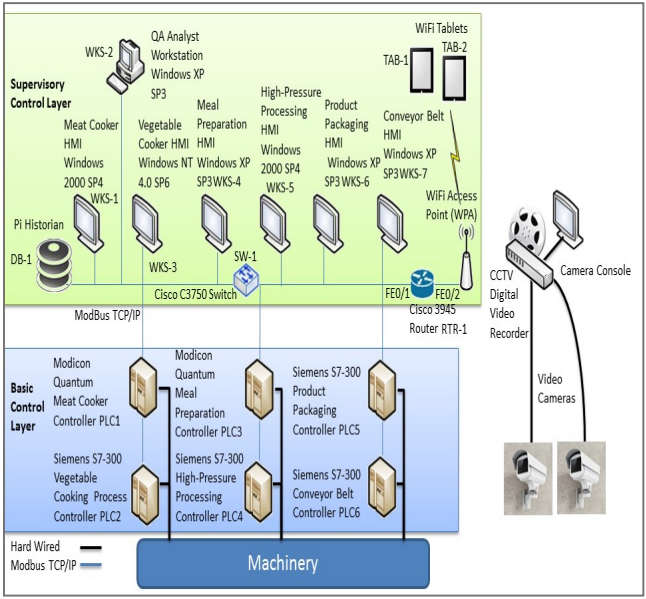

**Figure 5:** Plant Network used to control the AQUA Production Line

The plant network used by the AQUA is shown in Figure 5. It consists of six programmable logic controllers (PLCs), several workstations, a closed circuit television (CCTV) system, and a wireless network for tablet computers. Technicians use the tablet computers to access the human machine interface (HMI) displays. The plant network is not connected to the corporate network or the Internet. All plant machinery is hard-wired to the input and output modules of the PLCs. The CCTV cameras are hard-wired to the digital video recorder (DVR).

### C. Cyber Wargame Method and Execution

Before the wargame began, in addition to reviewing the information packet read-ahead document, both teams were allowed to ask the game facilitators specific technical questions. Answers were shared with both teams. During the exercise, the RED and BLUE teams met separately to review all of the technical information and discuss strategies. Each team documented their potential strategies. After this, the teams converged to share their findings. The RED team shared their attack strategies with the BLUE team. Mitigations, counter-attacks, and counter-mitigations were discussed. Details about the RED team strategies and BLUE team mitigations can be found in [20]. The representation of those results in terms of our game-theoretic framework (section II.A) is described in the following subsections.

### D. Description of Game-theoretic Variables

Attack and Defense Strategies

The RED team submitted five attack strategies $\{s_{a,i}\}, i = 1 \ldots 5$, which we list in **Table 1**.

| {i} | Attack Strategy $s_{a,i}$ | Description of Attack |
|---|---|---|
| 1 | Layer-2 Attack | First the RED team gains access by cracking the password of the wireless network. The authentication credentials of the wireless tablets are then spoofed to penetrate the rest of the plant network. Complete control of the PLC traffic is then gained. |
| 2 | Subversion and Espionage | The RED team coerces a plant employee to install a rogue wireless device or malware into the plant network. The RED team, via remote access to the wireless device or via malware, then misdirects commands to the PLCs. |
| 3 | Rival Employer Attack | Posing as a rival employer searching for new hires, the RED team acquires vital plant information by interviewing plant employees. When a job opening becomes available at the AQUA plant, an insider sponsored by the RED team takes the job and consequently performs malicious activity to gain access to the plant network and the PLCs. |
| 4 | Jumping the Airgap | The plant machine that writes the media for patching the plant equipment is attacked. Once that machine is owned, malware with capabilities for disrupting, destroying and Command and Control (C2) is installed and executed strategically. |
| 5 | Human Interface Device Attack | A human interface device (HID), such as a USB mouse, is infected with malware. It then launches additional malware into the plant network, ultimately disrupting or destroying the plant processes. |

**Table 1:** RED Team Attack Strategies.

The BLUE team proposed 15 specific mitigations (see Table *2*) for their defense against the proposed attack strategies. Since it is not feasible to consider all possible ($N_{d,max} = 2^{15} = 32,768$) defender strategies that could be constructed from the 15 mitigations, four defense strategies $\{s_{d,j}\}, j = 1 \ldots 4$ were identified by selecting specific mitigations. These four strategies offer potentially good security with reasonably acceptable costs. A fifth strategy "No Action" was also noted and is identified here as *j=0* for reference. Defender mitigation costs were estimated manually by the BLUE team and are also listed in the table. Complete details about the attack and defense strategies can be found in [20] and [22].

Defender Computations

We next describe our game-theoretical computations. Since all strategies by both BLUE and RED teams were known by both teams in the exercise, costs and penetration probabilities of the RED team were estimated based on the BLUE team mitigations. We therefore discuss the defender mitigations and computations first.

We first construct a translation matrix $M_{jm}$ between mitigations and defense strategies, based on BLUE team input shown in Table 2.

**Defense Mitigations and Strategies**

| {k} | Mitigation Description | Cost [k$] | No Action (j=0) | Basic Security (j=1) | IDS+ (j=2) | IDS Enhanced (j=3) | IDS+ and Physical Security (j=4) |
|---|---|---|---|---|---|---|---|
| 1 | Upgrade WiFi Security | 10 | | X | X | X | X |
| 2 | Install Hardware Firewalls | 15 | | | X | X | X |
| 3 | Install Network Honeypots | 30 | | | | | X |
| 4 | Configure VLANs | 4 | | | X | X | X |
| 5 | Install Clear Conduit | 8 | | | | X | |
| 6 | Safeguard Plant Documents | 10 | | X | X | X | X |
| 7 | Install IDS | 40 | | | X | X | X |
| 8 | Install Layer-2 IDS Sensor Feed | 40 | | | | | X |
| 9 | Apply STIG Controls | 25 | | X | X | X | X |
| 10 | Upgrade Training Methods | 15 | | X | X | X | X |
| 11 | Monitor Ports on Devices | 30 | | | | X | X |
| 12 | Disallow USB Media Installs | 5 | | X | X | X | X |
| 13 | Upgrade Scanning Station | 15 | | | X | X | X |
| 14 | Lock BIOS on Devices | 5 | | | X | X | X |
| 15 | Upgrade Physical Security for PLCs | 20 | | | | | X |
| | Defender Strategy Costs $C_{d,ij}$ [k$] (independent of attack {i}) | | 0 | 65 | 144 | 182 | 264 |

**Table 2:** BLUE Team Defense Strategies and Mitigations. Estimated costs are shown for individual mitigations and for the five strategies associated with the mitigations.

(5)

| {l} | Penetration Layer Description |
|---|---|
| 1 | Penetration into Wireless Layer-2 Network |
| 2 | Penetration of Router RTR-1 |
| 3 | Penetration of Switch SW-1 into Plant Network |
| 4 | Access to PLC |

**Table 3:** Penetration Layers for Attack

$$M_{jm} = \begin{bmatrix} 0 & 0 & 0 & 0 & 0 & 0 & 0 & 0 & 0 & 0 & 0 & 0 & 0 & 0 & 0 \\ 1 & 0 & 0 & 0 & 0 & 1 & 0 & 0 & 1 & 1 & 0 & 1 & 0 & 0 & 0 \\ 1 & 1 & 0 & 1 & 0 & 1 & 1 & 0 & 1 & 1 & 0 & 1 & 1 & 1 & 0 \\ 1 & 1 & 0 & 1 & 1 & 1 & 1 & 0 & 1 & 1 & 1 & 1 & 1 & 1 & 0 \\ 1 & 1 & 1 & 1 & 0 & 1 & 1 & 1 & 1 & 1 & 1 & 1 & 1 & 1 & 1 \end{bmatrix}.$$

The vector defining defender costs by mitigation $m$, $C_{d,m} =$
[10 15 30 4 8 10 40 40 25 15 30 5 15 5 20],
is also constructed (see **Table 2**). We use the translation matrix $M_{jm}$ to compute defender strategy costs as a function of defender $j$:

$$C_{d,j} = M_{jm} C_{d,m} = [0 \ \ 65 \ \ 144 \ \ 182 \ \ 264], \ j = 0...4,$$

where $M_{jm} C_{d,m}$ implies summation (matrix multiplication) over $m$. Note that defender costs are independent of attacker strategies $i$ in this table-top wargame example. This is a generalized case of our game-theoretic framework described in section II.A.

Although the defender costs are not dependent on the attacker strategies {i}, we construct the defender cost matrix over {ij}:

$$C_{d,ij} = \begin{bmatrix} 0 & 0 & 0 & 0 & 0 \\ 65 & 65 & 65 & 65 & 65 \\ 144 & 144 & 144 & 144 & 144 \\ 182 & 182 & 182 & 182 & 182 \\ 264 & 264 & 264 & 264 & 264 \end{bmatrix},$$

and use it to compute the defender utility. Here, we have assumed that the defender strives to keep the portion of the defender's assets $b$ that would have otherwise been forfeited to the attacker as a benefit. The probability that those assets are not forfeited is $p = 1 - P_{ij}^T$.

We can then compute the defender cost utility as:

$$u_{d,ij} = b(1 - P_{ij}^T) - C_{d,ij} = \begin{bmatrix} 0 & 235 & 226 & 188 & 657.25 \\ 0 & 435 & 776 & 798 & 734.75 \\ 0 & 935 & 856 & 818 & 736 \\ 500 & 810 & 743.5 & 705.5 & 623.5 \\ 500 & 685 & 606 & 568 & 486 \end{bmatrix},$$

where $P_{ij}^T$ is computed below.

Attacker Computations
Four attack layers (cf. Figure 3) were identified by the RED team. We reference the attack layers with index {l} and list descriptions of the layers in Table 3.

Attacker costs were estimated by the RED team for each attack strategy $s_{a,i}$. Costs were estimated generally as a function of BLUE team mitigation. This differs slightly from our mathematical framework which requires costs as a function of {ij}, so we must translate the RED team input appropriately. We separate the attacker costs according to whether the costs were dependent on defender strategies (mitigations) or not. Fixed costs are independent of mitigations and are listed in Table 4. Differential costs are dependent on mitigations and are listed in Table 5. Many of the fixed costs are associated with preliminary research, reconnaissance, and bribes which occur before the attack begins, i.e., before penetration of layer *1*.

**Fixed Attacker Costs and Success Probabilities**

| Attack {i} | Justification (for Costs and Probability of Success Estimations) | Layer {l} | Cost [hr or k$] | Probability of success | Mitigations of Interest |
|---|---|---|---|---|---|
| 1 | Research for WiFi attack | N/A | 120 hr | N/A | N/A |
| 1 | Time to brute-force carack WiFi password | 1 | 24 hr | 1.0 | 1 |
| 1 | Time to brute-force crack RTR-1 password | 2 | 48 hr | 1.0 | 8 |
| 1 | Time to brute-force crack SW-1 password | 3 | 48 hr | 1.0 | 8 |
| 1 | Time to set up modification of PLC traffic | 4 | 4 hr | 1.0 | 7,8,10 |
| 2 | Research, Bribe | N/A | $65k | 1.0 | N/A |
| 3 | Research, Bribe | N/A | $320k | 1.0 | N/A |
| 4 | Research, Set-up | N/A | $35k | 0.5 | N/A |
| 5 | Research, Bribe | N/A | $40k | 0.5 | N/A |

**Table 4:** Fixed Attacker Costs. These costs were independent of mitigation or defense strategy. We assume labor costs of 1k$ per hour of RED team activity.

For the purposes of computing our game-theoretic model, we construct attacker fixed cost $C^0_{a,i,j}$ and penetration probability $p^0_{i,j}$ matrices for attacker and defender strategies {i,j} using the fixed RED team input from Table 4:

$$C^0_{a,i,j} = \begin{bmatrix} 244 & 244 & 244 & 244 & 244 \\ 65 & 65 & 65 & 65 & 65 \\ 320 & 320 & 320 & 320 & 320 \\ 35 & 35 & 35 & 35 & 35 \\ 40 & 40 & 40 & 40 & 40 \end{bmatrix}, \text{ and}$$

$$p^0_{i,j} = \begin{bmatrix} 1.0 & 1.0 & 1.0 & 1.0 & 1.0 \\ 1.0 & 1.0 & 1.0 & 1.0 & 1.0 \\ 1.0 & 1.0 & 1.0 & 1.0 & 1.0 \\ 0.5 & 0.5 & 0.5 & 0.5 & 0.5 \\ 0.5 & 0.5 & 0.5 & 0.5 & 0.5 \end{bmatrix}.$$

**Differential Costs of Attacker**

| Attack {i} | Mitigation {m} | Justification (for Costs and Probability of Success Estimation) | Layer {l} | Cost [hr] | Probability of success |
|---|---|---|---|---|---|
| 1 | 1 | Additional time needed for cracking WiFi | 1 | 24 | 1.0 |
| | 7 | Higher probability of detection | 4 | 0 | 0.9 |
| | 8 | Higher probability of detection | 2 | 0 | 0.5 |
| | 8 | Higher probability of detection | 3 | 0 | 0.5 |
| | 8 | Higher probability of detection | 4 | 0 | 0.5 |
| | 10 | Advanced research needed and higher probability of detection | 4 | 4 | 0.7 |
| 2 | 2 | Lower probability of successful subversion | 4 | 0 | 0.2 |
| | 3 | Wastes labor and higher probability of detection | 4 | 20 | 0.7 |
| | 7 | Higher probability of detection | 4 | 0 | 0.8 |
| | 8 | Higher probability of detection | 2 | 0 | 0.5 |
| | 8 | Higher probability of detection | 3 | 0 | 0.5 |
| | 8 | Higher probability of detection | 4 | 0 | 0.5 |
| | 10 | Advanced research needed and higher probability of detection | 4 | 4 | 0.5 |
| | 11 | Higher probability of detection | 3 | 0 | 0.25 |
| 3 | 3 | Wastes labor and higher probability of detection | 4 | 20 | 0.7 |
| | 7 | Higher probability of detection | 4 | 0 | 0.9 |
| | 10 | Advanced research needed and higher probability of detection | 4 | 4 | 0.7 |
| | 12 | Defeats the attack | 3 | 0 | 0 |
| | 14 | Lower probability of success | 3 | 0 | 0.5 |
| 4 | 6 | Lower probability of success | 3 | 0 | 0.5 |
| | 10 | Advanced research needed and higher probability of detection | 4 | 4 | 0.5 |
| | 13 | Malware may be detected | 3 | 0 | 0.9 |
| 5 | 6 | Lower probability of HID installation | 3 | 0 | 0.5 |

**Table 5:** Differential Attacker Costs. These costs are dependent on BLUE team mitigations. We again assume labor costs of 1k$ per hour of RED team activity.

Next, we utilize the RED team input from Table *5* to construct differential attacker cost $C^i_{a,lm}$ and penetration probability $p^i_{a,lm}$ matrices, one for each attack {i}. Since the RED team cost definitions are dependent on BLUE team mitigations and penetration layer, we index these matrices over layer l and mitigation m. While this information could be stored and computed using 3-dimensional matrices $C_{a,ilm}$ and $p_{ilm}$, for the purposes of illustration, we use five separate 2-dimensional matrices (one for each attack $i$) in the following discussion.

For example, for attack 3, by examining Table 5, we can construct differential cost $C_{a,lm}^3$ and penetration probability matrices $p_{a,lm}^3$:

$$C_{a,lm}^3 = \begin{bmatrix} 0 & 0 & 0 & 0 & 0 & 0 & 0 & 0 & 0 & 0 & 0 & 0 & 0 & 0 \\ 0 & 0 & 0 & 0 & 0 & 0 & 0 & 0 & 0 & 0 & 0 & 0 & 0 & 0 \\ 0 & 0 & 0 & 0 & 0 & 0 & 0 & 0 & 0 & 0 & 0 & 0 & 0 & 0 \\ 0 & 0 & 20 & 0 & 0 & 0 & 0 & 0 & 0 & 4 & 0 & 0 & 0 & 0 \end{bmatrix}, \text{ and}$$

$$p_{a,lm}^3 = \begin{bmatrix} 1 & 1 & 1 & 1 & 1 & 1 & 1 & 1 & 1 & 1 & 1 & 1 & 1 & 1 \\ 1 & 1 & 1 & 1 & 1 & 1 & 1 & 1 & 1 & 1 & 1 & 1 & 1 & 1 \\ 1 & 1 & 1 & 1 & 1 & 1 & 1 & 1 & 1 & 1 & 0.0 & 1 & 0.5 & 1 \\ 1 & 1 & 0.7 & 1 & 1 & 1 & 0.9 & 1 & 1 & 0.7 & 1 & 1 & 1 & 1 \end{bmatrix}.$$

Since our game-theoretical model references defender strategy $j$ instead of mitigation $m$, we recompute the differential attacker cost over $\{l,j\}$: $C_{a,lj}^3 = C_{a,lm}^3 M_{mj}$, where $M_{mj} = M_{jm}^T$ is the transpose of $M_{mj}$. The result for attack $i=3$ is shown below.

$$C_{a,lj}^3 = \begin{bmatrix} 0 & 0 & 0 & 0 & 0 \\ 0 & 0 & 0 & 0 & 0 \\ 0 & 0 & 0 & 0 & 0 \\ 0 & 4 & 4 & 4 & 24 \end{bmatrix}$$

We next need to condense $\{l\}$ and construct a single differential cost matrix for all attacks $\{i\}$. By summing row vectors over $\{l\}$, we obtain $C_{a,j}^3 = [0\ 4\ 4\ 4\ 24]$ for attack 3, and perform similar calculations for the other four attacks. The differential cost matrix $C_{a,ij}^D$ over $\{i,j\}$ is then:

$$C_{a,ij}^D = [C_{a,j}^1; C_{a,j}^2; C_{a,j}^3; C_{a,j}^4; C_{a,j}^5] = \begin{bmatrix} 0 & 28 & 28 & 28 & 28 \\ 0 & 4 & 4 & 4 & 4 \\ 0 & 4 & 4 & 4 & 24 \\ 0 & 4 & 4 & 4 & 4 \\ 0 & 0 & 0 & 0 & 0 \end{bmatrix}.$$

We can now compute the total attacker cost matrix $C_{a,ij}$ over $\{i,j\}$:

$$C_{a,ij} = C_{a,ij}^0 + C_{a,ij}^D = \begin{bmatrix} 244 & 272 & 272 & 272 & 272 \\ 65 & 69 & 69 & 69 & 69 \\ 320 & 324 & 324 & 324 & 344 \\ 35 & 39 & 39 & 39 & 39 \\ 40 & 40 & 40 & 40 & 40 \end{bmatrix}.$$

The differential penetration probabilities for the mitigations are also computed from information provided by the attacker (given in Table 5) by constructing the differential penetration probabilities $p_{im}^D$ for each attack $\{i\}$ and each mitigation $\{m\}$:

$$p_{im}^D = \begin{bmatrix} 1 & 1 & 1 & 1 & 1 & 1 & 0.9 & 0.125 & 1 & 0.7 & 1 & 1 & 1 & 1 \\ 1 & 0.2 & 0.5 & 1 & 1 & 1 & 0.8 & 0.125 & 1 & 0.5 & 0.25 & 1 & 1 & 1 \\ 1 & 1 & 0.7 & 1 & 1 & 1 & 0.9 & 1 & 1 & 0.7 & 1 & 0.0 & 1 & 0.5 & 1 \\ 1 & 1 & 1 & 1 & 1 & 0.5 & 1 & 1 & 1 & 0.5 & 1 & 1 & 0.9 & 1 & 1 \\ 1 & 1 & 1 & 1 & 1 & 0.5 & 1 & 1 & 1 & 1 & 1 & 1 & 1 & 1 \end{bmatrix}.$$

Each entry is computed by multiplying probabilities over layers $\{l\}$ for each $\{i,m\}$ combination. For example, from Table 5, attack 2, mitigation 8 has penetration probabilities of 0.5 for each of layers 2, 3, and 5, so that $p_{im}^D = (0.5)^3 = 0.125$.

Once $p_{im}^D$ is constructed, we then use the translation matrix $M_{mj}$ to calculate and redistribute the probabilities as a function of defender strategy $j$. For example, $p_{ij}$ for attack $i=3$ and defense $j=2$ includes mitigations 1, 6, 9, 10, and 12, which have probabilities of success of 1.0, 1.0, 1.0, 0.7, and 0.0, respectively. These five values multiply to $p_{ij=32}^D = 0.0$. Continuing this process, we find the resulting differential probabilities of success for all layers $\{l\}$, as a function of attack and defense strategies $\{i,j\}$ is:

$$p_{ij}^D = \begin{bmatrix} 1 & 0.7 & 0.63 & 0.63 & 0.07875 \\ 1 & 0.5 & 0.08 & 0.02 & 0.00125 \\ 1 & 0.0 & 0.0 & 0.0 & 0.0 \\ 1 & 0.25 & 0.225 & 0.225 & 0.225 \\ 1 & 0.5 & 0.5 & 0.5 & 0.5 \end{bmatrix}.$$

The total probability matrix can then be computed by taking the element-by-element product with the fixed probability matrix: $P_{T,ij} = p_{ij}^0 p_{ij}^D$:

$$P_{ij}^T = \begin{bmatrix} 1 & 0.7 & 0.63 & 0.63 & 0.07875 \\ 1 & 0.5 & 0.08 & 0.02 & 0.00125 \\ 1 & 0.0 & 0.0 & 0.0 & 0.0 \\ 0.5 & 0.125 & 0.1125 & 0.1125 & 0.1125 \\ 0.5 & 0.25 & 0.25 & 0.25 & 0.25 \end{bmatrix}.$$

The utility cost of the attacker $u_{a,ij}$ is computed from equation (1b). For our table-top wargame, the benefit $b$ was $100k, so that:

| Defender Cost Utility [k$] | | | | | |
|---|---|---|---|---|---|
| | Defense Strategy {j} | | | | |
| | j=0 | j=1 | j=2 | j=3 | j=4 |
| Attack Strategy {i} i=1 | 0 | 235 | 226 | 188 | 657.25 |
| i=2 | 0 | 435 | 776 | 798 | 734.75 |
| i=3 | 0 | 935 | 856 | 818 | 736 |
| i=4 | 500 | 810 | 743.5 | 705.5 | 623.5 |
| i=5 | 500 | 685 | 606 | 568 | 486 |

**Table 6:** Defender Cost Utility $u_{d,ij}$ for attack and defense strategies {i,j} computed assuming the BLUE team retains the value $b$ of their assets with probability $1 - P_{T,ij}$.

$$u_{a,ij} = b P_{ij}^T - C_{a,ij}^T =$$

$$\begin{bmatrix} 756 & 428 & 358 & 358 & -193.25 \\ 935 & 431 & 11 & -49 & -67.75 \\ 680 & -324 & -324 & -324 & -344 \\ 465 & 86 & 73.5 & 73.5 & 73.5 \\ 460 & 210 & 210 & 210 & 210 \end{bmatrix}$$

## IV. DISCUSSION

As described in section II, the objective of our game-theoretic model is to provide a useful analytical tool for the attacker and defender to select their strategies based on cost utilities (Equations 1 and 2) and penetration probability (Equation 3).

| Attacker Cost Utility [k$] | | | | | |
|---|---|---|---|---|---|
| | Defense Strategy {j} | | | | |
| | j=0 | j=1 | j=2 | j=3 | j=4 |
| Attack Strategy {i} i=1 | 756 | 428 | 358 | 358 | -193.25 |
| i=2 | 935 | 431 | 11 | -49 | -67.75 |
| i=3 | 680 | -324 | -324 | -324 | -344 |
| i=4 | 465 | 86 | 73.5 | 73.5 | 73.5 |
| i=5 | 460 | 210 | 210 | 210 | 210 |

**Table 7:** Attacker Cost Utility $u_{a,ij}$ for attack and defense strategies {i,j} computed from RED team input from Table 4 and Table 5

For convenience, we provide the attacker and defender cost utilities (payoff matrices) and total penetration probability matrix in Table 7, Table 6 and Table 8, respectively.

As mentioned, in both our table-top wargame and our game-theoretic framework (section II), we assume that defender and attacker have full knowledge of all player strategies and costs, and are able to compute the utility and penetration probability matrices.

| Total Attack Penetration Probabilities | | | | | |
|---|---|---|---|---|---|
| | Defense Strategy {j} | | | | |
| | j=0 | j=1 | j=2 | j=3 | j=4 |
| Attack Strategy {i} i=1 | 1 | 0.7 | 0.63 | 0.63 | 0.08 |
| i=2 | 1 | 0.5 | 0.08 | 0.02 | 0.001 |
| i=3 | 1 | 0.0 | 0.0 | 0.0 | 0.0 |
| i=4 | 0.5 | 0.125 | 0.11 | 0.11 | 0.11 |
| i=5 | 0.5 | 0.25 | 0.25 | 0.25 | 0.25 |

**Table 8:** Attack Penetration Probabilities $P_{T,ij}$ for attack and defense strategies {i,j} computed from RED team input from Table 4 and Table 5

Given all this information, how should players select the most advantageous strategies? Earlier in sections I.B and II.B, we discussed general approaches to such a selection. In the following sections, we consider how to apply those approaches in our example wargame.

| Preferred Cost Utility Strategies | | | | | |
|---|---|---|---|---|---|
| | Defense Strategy {j} | | | | |
| | j=0 | j=1 | j=2 | j=3 | j=4 |
| Attack Strategy {i} i=1 | | | A | A | D |
| i=2 | A | A | | D | |
| i=3 | | D | | | |
| i=4 | | D | | | |
| i=5 | | D | | | A |

**Table 9:** Preferred cost utility strategies (see Equations 4a and 4b) for the attacker (A) and the defender (D).

### A. Pure Strategy Equilibrium

If both players select their strategies simultaneously and each act to maximize their own utility, an equilibrium strategy pair will exist if the conditions outlined in section II.B lead to a single unique strategy pair. That is, there is a strategy pair $\{i^*, j^*\}$ that is a preferred ("most likely") strategy for both players. To translate this into a military analogy, a single unique strategy pair might mean that the defender prefers to defend at the same position where the attacker prefers to attack. In the shorthand notation of Equations (4a) and (4b), $i^* = \{i_j^*\}$ and $j^* = \{j_i^*\}$ are unique. For our table-top exercise, the preferred strategies are shown in Table 9. These values are found manually by identifying the maximum attacker cost utility across the columns in Table 7 and the maximum defender utility across the rows in Table 8. Preferred defender strategies are marked with an "A" and preferred attacker strategies are marked with a "D." As one can see, there is no single {ij} pair such that both i and j are preferred strategies, thus there is no equilibrium solution. To rephrase, there is no such a case here that the defender and the attacker prefer the same combination of their respective strategies.

However, consider a situation where players are uncertain about cost and benefit estimates – a likely situation in the domain of cybersecurity. The players may decide instead to use the penetration probabilities alone as utility functions so that the attacker aims to maximize $P_T$ and the defender aims to minimize the same value. That is, they intend to choose the most or least probably success of acquiring the target regardless of the cost. Their respective preferred strategies in this case are shown in Table 10. Here the maximum across rows and columns is calculated in the same manner using the probability penetration matrix (Table 8). Note that multiple maxima and minima are found since some cells have the same value across the rows or columns. In this case, there is a single common preferred strategy pair {5,4} that will satisfy both players as an equilibrium point, so that if each player used this method to make a decision, no alternative strategy would give any players an advantage. To rephrase, in this case the attacker is satisfied that that the defender picks the strategy 4, and the defender is satisfied that the attacker picks the strategy 5.

### B. Considering Multiple Strategies of the Opponent

Although we are not exploring here a mixed-strategy approach, such as the Strong Stackelberg Strategy Equilibria mentioned earlier, we can consider multiple strategies in our example wargame. For the sake of concreteness, let's adopt the perspective of the BLUE team, the defenders. Looking at Table 7, the BLUE team would notice that from the attacker's perspective strategy i=5 is generally better than i=4. Indeed, with i=5, the attackers gains better utility for all defense strategies except j=0, and even in the case of j=0, the benefit of i=4 for the attacker is insignificant (460 vs 465). Thus, if the cost utility is the primary consideration for the attacker, the attacker would surely select i=5 over i=4. Similar considerations may apply when the BLUE team explores the attacker's strategies i=1, 2 and 3. After consideration, the BLUE team might conclude that it needs to select a strategy that defends the best against both strategies i=1 and i=5. Then, looking at Table 6, the BLUE team may decide that the defense strategy j=4 offers the BLUE team a relatively safe defense (in terms of cost utility) against both i=1 and i=5. Thus, the BLUE team could select j=4 as its preferred strategy in this manner based on the cost-utility matrices. To rephrase, here the defender decided that the most likely strategies of the attacker are the strategy 5 and the strategy 1, and then selected its own strategy 4 as it provides better outcomes whether the attacker chooses 5 or 1.

### C. Playing Against the Most Likely Strategy of the Opponent

Suppose the BLUE team has additional information that the attacker tends to be risk-averse. In that case, the attacker is likely to prefer i=5 over i=1, because i=1 might cause the attacker heavy loss (-193.25 in case the defender selects j=4). With this, the BLUE team can conclude that the attacker has a single "most likely" strategy i=5. In that case, the BLUE team selects defense strategy j=1 (with its defender's cost utility of 685) as its best play against the attacker's i=5. To rephrase, here the defender decided that the one most likely strategy of the attacker is the strategy 5, and then selected its own strategy 4 as the most likely to bring the best outcome against the attacker's strategy 5.

| **Preferred Penetration Utility Strategies** | | | | | | |
|---|---|---|---|---|---|---|
| | | **Defense Strategy {j}** | | | | |
| | | j=0 | j=1 | j=2 | j=3 | j=4 |
| **Attack Strategy {i}** | i=1 | A | A | A | A | D |
| | i=2 | A | | | | D |
| | i=3 | A | D | D | D | D |
| | i=4 | | | D | D | D |
| | i=5 | | D | D | D | D,A |

**Table 10:** Preferred strategies for the attacker (A) and the defender (D) when the penetration probability matrix is used as a utility function instead of the cost utility matrices.

### D. Playing Against the Most Damaging Strategy of the Opponent

Alternatively, the BLUE team might consider on how to avoid heavy losses, rather than how to gain best cost utility. Looking at Table 6, the BLUE team would notice the attacker strategy i=1 brings the defender the least utility against 4 out of 5 defense strategies. In other words, the attacker strategy i=1 is the "most damaging" to the defender. In this case, j=4 could be a strong strategy for the defender. However, considering that the "most likely" strategy of the attacker is i=5 (as we determined above), it is safer for the BLUE team to select j=1, which gives the BLUE team a relatively good cost utility of 235 even if the attacker selects the "most damaging" i=1. To rephrase, here the defender decides to focus on two strategies of the attacker – the most likely strategy 5 and the most damaging strategy 1 – and then selects its own strategy 1 that offers acceptable losses regardless of whether the attacker chooses the attacker's strategy 5 or 1.

### E. Consequential Moves

In summary, our game-theoretic model computations outlined in section II.A offer a set of tools in the form of matrices that the wargame players may use for selecting their initial strategies. While we do not describe consequential strategies in this work, the model could be adapted so that similar matrix calculations could be used to consider consequential strategies. Both teams would need to reassess new costs and penetration probabilities given their current position in the game. An elaborate game-in-game theoretical framework for solving a complete path through the layers of defense is described in [23]. In implementing such a decision scheme in practice, one must incorporate the effects that increasingly accurate information of the attacker has on his consequential moves as he penetrates the layers.

In the case that either the defender or the attacker selects his strategy first and the opponent follows, any of the SSE computational methods may be used to find a vector of weights for the leader initial strategy [19]. The follower can then use the payoff and penetration probability matrix to decide on his next strategy.

## V. Conclusions

While our methods may have limitations in real-world scenarios, the game-theoretic model we outlined in this paper is suitable for assisting wargaming teams. We demonstrate an application of our model to a fairly realistic wargame for which opposing teams have complete knowledge of the system and each others' strategies.. Specifically, we find that the use of the model benefits the participants of a cyber wargame in several ways:

### A. Elucidation of costs, benefits and assumptions

The players – either within a single team or in joint discussions by both BLUE and RED teams – are motivated to explicitly define, quantify and document a number of critical elements of their decision making process. They discuss and document layers of the defense that the attacker must penetrate. They think about costs and benefits from the perspective of both teams. They are compelled to search for additional pertinent information about costs and timing of attacks and of defensive mitigations, as well as their probabilities of success. They uncover and clarify inconsistent assumptions. Ultimately, they either arrive to a well-reasoned consensus or to a clear understanding of where the opinions differ. All these considerations and conclusions are likely to be well-documented as a result of these steps in the wargaming process

### B. Enhancement of potential strategies

Having distilled their assumptions and estimates into overall quantifications of utilities and probabilities (such as Table 7, Table 6, and Table 8), the BLUE and RED teams can examine the data for the purposes of identifies opportunities for strategy improvements. For example, the RED team might notice that attack strategy i=4 is clearly inferior to i=5, from the perspective of the attacker cost utility. Should the strategy i=4 be eliminated from the attacker's repertoire? Or, is this an indication of incomplete or inaccurate information somewhere in the process? Or, could the strategy i=4 be improved in a way that would make it competitive with i=5? These are questions that are not likely to be asked – or answered -- without such a quantitative analysis.

### C. Aticulated selection of strategies

The players can use this analytical process to engage in rational assessment of alternatives, and arrive to a selection of an appropriate strategy in a well-reasoned manner. The reasons for preferring a certain strategy can be clearly stated, often in a quantitative manner, and underlying reasons can be readily explained. Without such an analytical tool, cyber wargamers often have difficulties making a rational, explainable strategy selection.

### D. Future Research

While this line of research can be extended in a number of directions, one of them is particularly salient: relaxing the assumption that both the attacker and the defender have full knowledge of the systems and technologies available to both sides. Relaxation of this simplifying assumption raises numerous challenging research questions. For example, how to make appropriate assumptions about the opponent's lack of knowledge? How does making such an assumption increases the risk that the selected strategy might fail if the assumption is wrong? How can such relaxation of assumptions be portrayed realistically in a human wargaming?

## Acknowledgments

We would like to thank Prof. Quanyan Zhu and Dr. Jeffrey Pawlick for thoughtful discussions and suggestions.

The views expressed in this article are those of the authors and do not reflect the official policy or position of the US Army, Department of Defense, or the US Government.